# A WAV2VEC2-BASED EXPERIMENTAL STUDY ON SELF-SUPERVISED LEARNING METHODS TO IMPROVE CHILD SPEECH RECOGNITION.

**Rishabh Jain[1], (Graduate Student Member, IEEE), Andrei Barcovschi[1], Mariam Yiwere[1], Dan Bigioi[1] (Graduate Student Member, IEEE), Peter Corcoran[1], (Fellow, IEEE) and Horia Cucu[2], (Member, IEEE)**
[1] School of Electrical and Electronics Engineering, University of Galway, Galway, H91 TK33 Ireland
[2] Speech and Dialogue Research Laboratory, University Politehnica of Bucharest, Romania

Corresponding author: Rishabh Jain (e-mail: rishabh.jain@universityofgalway.ie)

**ABSTRACT** Despite recent advancements in deep learning technologies, Child Speech Recognition remains a challenging task. Current Automatic Speech Recognition (ASR) models require substantial amounts of annotated data for training, which is scarce. In this work, we explore using the ASR model, wav2vec2, with different pretraining and finetuning configurations for self-supervised learning (SSL) toward improving automatic child speech recognition. The pretrained wav2vec2 models were finetuned using different amounts of child speech training data, adult speech data, and a combination of both, to discover the optimum amount of data required to finetune the model for the task of child ASR. Our trained model achieves the best Word Error Rate (WER) of 7.42 on the MyST child speech dataset, 2.91 on the PFSTAR dataset and 12.77 on the CMU KIDS dataset as compared to any other previous methods. Our models outperformed the wav2vec2 BASE 960 on child speech which is considered a state-of-the-art ASR model on adult speech by just using 10 hours of child speech data in finetuning. The analysis of different types of training data and their effect on inference is also provided by using a combination of datasets in pretraining, finetuning and inference.

**INDEX TERMS** Child speech recognition, Self-supervised learning, wav2vec2, Automatic Speech Recognition, MyST Dataset, PFSTAR Dataset, CMU_Kids Dataset

## I. INTRODUCTION

Current deep learning-based automatic speech recognition models perform remarkably well on adult speech data. However, they struggle when it comes to recognizing speech from children. Models such as wav2vec2, Deep Speech 2, ContextNet, and others [1]–[7] all achieve impressive results on adult speech datasets such as LibriSpeech (~1000h), TIMIT (5.4h), LJSpeech (~24h), MediaSpeech (~10h), and more. This is due in no small part to the vast amounts of annotated adult speech data available for training such models and the ease with which it can be obtained. However, when it comes to child speech recognition, State-Of-The-Art (SOTA) ASR models trained on adult data perform quite poorly on child voice datasets. This is due to the inherent differences between adult and children's voices. A child's voice is quite different from an adult's voice [8], [9] in terms of pitch, linguistic and acoustic features, ability to understand and pronounce words, high fundamental frequency, and shorter vocal tract length. In addition, it is a challenging task to collect and annotate child speech data in comparison to adult speech data which can be acquired from various sources such as movies, news broadcasts, audiobooks, the internet, etc. Even if child speech can be collected from such sources, providing accurate annotations remains challenging. When compared to adult voice datasets, child voice datasets are quite limited [10].

### A. RELATED WORKS

In the past few years, there have been many different approaches to improving the performance of automatic child speech recognition systems [11]. Most of these approaches consist of various data augmentation techniques for increasing the amount of usable training data. Text-to-Speech based data augmentations as introduced by [12], [13], where ASR models are finetuned using synthetic data, have not shown significant increases in the accuracy of child ASR. Generative Adversarial Network (GAN) based augmentation [14]–[16] has also been explored to increase the amount of labeled data with acoustic attributes like those of child speech. Some of the other popular augmentation approaches include Vocal Tract Length Perturbation [17], Fundamental frequency feature normalization [18], out-of-domain data augmentation using Stochastic Feature Mapping (SFM) [19], and data processing-based augmentations [20] such as Speed Perturbation, Pitch Perturbation, Tempo Perturbation, Volume Perturbation, Reverberation Perturbation, and Spectral Perturbation. Spectrogram Augmentation also seems promising for improving the performance of ASR systems [21], [22]. Each of these methods shows improvements in child ASR accuracy, however, they still require corresponding labeled annotations to speech data.

Another recent trend is the use of transfer learning approaches for improving the recognition in child ASR for features adaptablity from adult to child speech. The authors in [23] perform extensive analysis to understand the effect of the amount of adaptation data, different Deep Neural Network (DNN) transfer learning configurations, and their impact on different age groups for improving child ASR. In [24], the authors explored the use of a two-step training strategy, which involves multilingual pretraining followed by transfer learning, for improving the performance of ASR systems on child speech.

Each of these methods show some improvements in child ASR accuracy, however, they still require corresponding labeled annotations to speech. A recent review of child ASRs [21] determined that most of these SOTA methods are supervised learning approaches. The authors in [25] show the performance of various supervised learning approaches for ASR in child speech. They compared the performance of end-to-end ASR systems with that of Deep Neural Network-Hidden Markov Model (DNN-HMM) hybrid systems. Another paper [26] studied the performance of Factored Time Delay Neural Networks (TDNN-F) with traditional and SOTA systems for ASR of child speech. These supervised approaches rely on labeled child speech data during training for the task of ASR.

As there is a distinct lack of labeled child speech data compared to adult, approaches that utilize unsupervised [27] and self-supervised learning [1] were explored for this paper. Therefore, the goal of this work is to present a method to incorporate unlabeled child speech data into the training procedure of a typical ASR model while also making use of abundant, labelled, and unlabeled adult speech data to improve the overall accuracy of ASR models on child speech.

## B. SELF-SUPERVISED LEARNING FOR CHILD ASR
Self-supervised learning (SSL) has emerged as a paradigm to learn general data representations from substantial amounts of unlabeled examples allowing one to then fine-tune models on small amounts of labeled data. The use of SSL for child ASR was first seen at Interspeech2021, where a model using SSL [28] received first place. A similar use case [20] was also presented in the SLT 2021 Children Speech Recognition challenge [29]. Another approach is used in [30], where the author uses a bidirectional unsupervised model pretraining with child speech ASR. After reviewing various approaches to SSL, wav2vec2 [1] was chosen for this paper. Wav2vec2 shows that using self-supervised learning for the task of ASR provides improvements over SOTA supervised learning approaches.

At the time of working on this paper, many applications of the wav2vec2 model for child speech recognition were observed. The authors in [31] propose the use of a transformer model pretrained on adult speech to achieve SOTA results on children's dataset. [32] provides a comparison between different SSL approaches for child speech recognition tasks. In [33], authors proposed a Domain Responsible Adaptation and Fine-Tuning (DRAFT) framework to address the domain shift between adult speech data used for pretraining and child speech data used for fine-tuning. They use wav2vec2 along with other SSL methods to examine the cross-domain transfer between different children's datasets.

This paper explores various pretraining and finetuning configurations with different combinations of adult and child speech datasets using wav2vec2 speech representations. Three child speech datasets were used in this study. These datasets were cleaned and preprocessed to make them usable for ASR. We also report the best results on different child speech validation. The ideal data requirement for pretraining and finetuning in a low-data scenario was also explored in this paper by observing the relation/pattern of performance in different datasets used.

The rest of this paper is organized as follows: Section 2 describes the model architecture. Section 3 introduces the datasets used for this paper. Section 4 includes the experiments and results. Conclusions are presented in Section 5.

## II. TRAINING METHODOLOGY FOR SSL
The wav2vec2 model [1] is used to extract speech representations from raw audio files in a self-supervised learning scenario and use these representations for ASR-specific tasks. Wav2vec2 is used in this paper as it can achieve state-of-the-art results when trained on a large amount of unlabeled speech data and finetuned on labeled data as small as 10 minutes. This is ideal for our task, as it is much easier to obtain significant amounts of unlabeled child speech data than gather accurately labeled data.

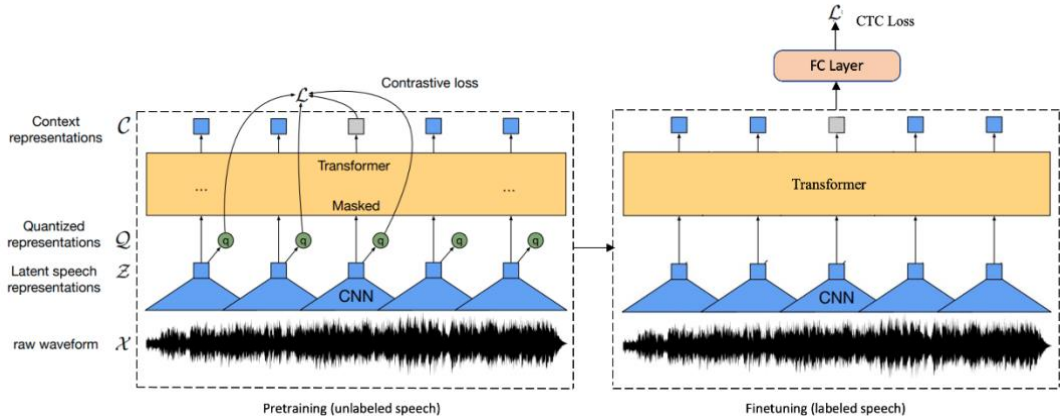

**FIGURE 1.** PRETRAINING AND FINETUNING STEPS IN WAV2VEC2 (FROM [1])

As it is a two-step training method, the first step includes a pretraining step in which the model is trained with a large amount of unlabeled data. Wav2vec2 uses a multilayer Convolutional Neural Network (CNN) to encode the speech audio. After encoding, masking is applied to the resulting latent speech representation which is fed into a transformer network to build a contextualized representation of the speech audio. Gumbel softmax [34] is used to calculate the contrastive loss on which the model is trained. Speech representations are learned from this training.

The second step includes finetuning on labeled data using Connectionist Temporal Classification (CTC) loss [35] for downstream ASR tasks. As the model learns SSL speech representation in pretraining, it can be trained using large quantities of unlabeled speech data and can be finetuned with only a small amount of labeled data. This way, the problem of scarcity of child speech is solved as we can train the 'pretraining' model with a combination of unlabeled speech data and it can also be used to learn speech representations from adult speech datasets making use of the abundant adult speech data.

### A. PRETRAINING
The pretraining stage of the wav2vec2 model consists of a feature encoder, context network, and quantization module. A feature encoder encodes the raw audio files with temporal convolution followed by batch normalization which is further normalized to zero mean and unit variance. A context network takes in the output of the feature encoder to calculate relative position embeddings. The quantization module is then used to calculate the contextualized representation by using Gumbel softmax and choosing the quantized representation from multiple codebooks and concatenating them. Experiments' configurations are provided as the BASE and LARGE models. The configurations differ in transformer block size but use the same size for the encoder. The feature encoder contains seven blocks with each block having strides of (5,2,2,2,2,2,2) and kernel widths of (10,3,3,3,3,2,2) and output temporal convolution of 512 channels. The context network of the BASE model contains 12 transformer blocks, each block with a 512-dim model, 8 attention heads, and a 2048-dim feed-forward inner layer, while the LARGE model contains 24 transformer blocks with model dimensions 1024, inner dimensions 4096, and 16 attention heads.

We use 4 NVIDIA Tesla V100 GPUs to pretrain the model. Model pretraining was optimized using ADAM [36]. During the first 8% of updates, the learning rate warms up to a peak of $5 \times 10^{-4}$ for BASE and $3 \times 10^{-4}$ for LARGE, and then it linearly decays. We use both BASE and LARGE models according to dataset size used for pretraining.

### B. FINETUNING
For finetuning, 29 tokens were used (from the Librispeech dataset) as provided by the authors in wav2vec2[1]. Models are optimized by minimizing CTC loss [35]. A modified version of SpecAugment [19] is applied as masking to timestamps and channels to reduce the overfitting and improve the recognition robustness. We fine-tune on one V100 GPU. For the first 1000 updates, only the final output classifier was trained, after which the Transformer block was also trained. The feature encoder was frozen during finetuning training. We also use different finetuning configurations depending on the size of finetuning datasets. The hyperparameters are kept the same as provided by the wav2vec2 authors [1]. The learning rate changes according to the dataset size as documented by the authors of wav2vec2 [1].

As the goal of this study is to evaluate the performance of self-supervised speech representations, it was decided not to incorporate a language model in this research. Additionally, previous research has shown that the best results for children's ASR systems were achieved without the use of an external language model [25]. Language model adaptation for child speech is also an unexplored research area. Child speech would require a specialized trained language model for best results. As there isn't any definitive publicly available language model for child speech, we consider this as a part of the future research topic.

## III. DATASET DESCRIPTION AND USAGE

The datasets are divided according to their usage. The child speech data used in this paper include MyST Child Corpus [37], CMU_Kids [38]and PF-STAR [39]. Adult Speech datasets include Librilight [40], LibriTTS[41], and LibriSpeech [42]. Please see Table 1 for details on the datasets' usage.

### A. DATASET DESCRIPTION

Below we provide a description of the datasets used in this paper.

#### 1) LIBRISPEECH [42]

Librispeech is an adult speech dataset with approximately 1000 hours of recorded audio with a sampling rate of 16Khz. The data is derived from read audiobooks from the LibriVox project. The data is carefully segmented, aligned, and used popularly in speech research.

#### 2) LIBRILIGHT [40]

Librilight is an adult speech dataset used as a benchmark for training speech recognition systems with limited or no supervision. It contains 60,000 hours of unlabeled adult speech extracted from audiobooks. It was mentioned in the wav2vec2 paper [1] and used by the authors.

#### 3) LIBRITTS [41]

The LibriTTS dataset is a large-scale dataset for training Text-To-Speech (TTS) models and is a subset of the Librispeech dataset. It consists of approximately 560 hours of high-quality audio and text transcriptions from audiobooks. This dataset is used here for inference over adult speech as it is a clean and noise-free dataset. It should help understand the relationship between child and adult speech datasets over inference. The 'dev-clean' segment of the LibriTTS dataset is used in this paper as it contains over 8.9 hours of clean adult speech. It is also widely used as a baseline in the validation of ASR and TTS experiments.

#### 4) MY SCIENCE TUTOR (MYST) CHILD SPEECH [37]

The MyST (My Science Tutor) Children's Speech Corpus consists of 393 hours of American English children's speech with a total of 228,874 utterances. The speech was collected from 1371 third, fourth and fifth-grade students. 45% of the utterances have been transcribed at the word level amounting to 197 hours. This dataset is used in this paper as it's the largest corpus of child speech available open source for research use. Even though data is 'noisy' (More details in B. Dataset Cleaning and Processing), its use can be seen in recent child speech research.

#### 5) PF-STAR CORPUS OF BRITISH ENGLISH CHILD SPEECH [39]

This corpus contains British English child speech from 158 children aged 4 to 14 years. The recordings are divided into a training set (7.5 hours), an evaluation set (1 hour) and a test set (5.6 hours). The corpus was collected at three locations: a university laboratory and two primary schools. It contains both read and spontaneous child speech with transcriptions. This dataset is also widely used in child speech ASR research.

#### 6) CMU KIDS [38]

CMU KIDS Corpus contains read-aloud sentences by children. It was created to provide training data for the SPHINX II automatic speech recognizer at Carnegie Mellon University. It contains 9 hours of American English child speech. The dataset contains 24 male and 52 female speakers having a total of 5180 utterances. It is also a widely used dataset for child ASR research. More details are in the next section.

### B. DATASET CLEANING AND PROCESSING

All speech data was converted into a 16-bit mono channel with a 16Khz sampling rate, wherever required. All the transcriptions were cleaned and normalized to remove abbreviations, punctuations, whitespaces, etc. and all the characters were changed to uppercase. All the non-linguistic annotation symbols (in child speech datasets) such as "<unk>, sil, hmm, <breath>, <noise>, <indiscernible>, [ze-], [cham-], [***ision], etc." were removed and only alphanumeric characters were retained in the transcript. This was done for all the labeled data used in this paper. Child datasets required further cleaning and pre-processing as follows:

#### 1) MYST CLEANUP

The MyST dataset contains over 393 hours of speech data, with only 197 hours of transcribed speech data presented in .trn file format. For the finetuning setup with child speech, the 10-20 second long speech samples from transcribed MyST were selected. Within the MyST dataset, samples below 10 seconds in length generally contained non-meaningful, noisy speech, and data above 20 seconds would lead to the GPU running out of memory. The MyST dataset contained a lot of noisy and non-meaningful sentences such as:

- o  <silence> I'm i don't know <noise> actually
- o  <whisper> sending go back (*)
- o  <whisper> what's this one <side_speech> it's an instinctive
- o  give me that <indiscernible> a circuit is a pathway
- o  <laugh> yeah yeah

The content between '<' and '>' tags were removed from all the transcriptions along with the tags themselves. All the cleaned text files were saved in a .txt format. A few other examples of MyST sentences can be seen in Figure 2.

```
009.wav|they're from different places
011.wav|the delta has more um clay (()) has more gravel mountains have more
015.wav|um
017.wav|um the bottom layer rocky the middle layer is more like clay and silt
020.wav|<no_signal>
021.wav|<no_signal>
023.wav|(())
002.wav|soils and rocks and land forms and stuff like that
005.wav|weathering or i think it was
```

**FIGURE 2. Example of transcript content containing non-meaningful annotations in the MyST dataset**

These utterances were further cleaned by manually listening and going through the transcripts, which amounted to a total of 65 hours of clean data. The data was then randomly split into two groups having 55 hours of data for training and 10 hours for testing as can be seen in Table 1.

## 2) PFSTAR CLEANUP

The PFSTAR corpus also contained a lot of non-meaningful utterances and noisy data samples. The dataset comes with '.trs' transcription files, containing time-aligned text information (see Figure 3). These timestamps were used to further segment the data into small audio chunks and remove noise from the dataset. The 'sp' tag from the transcription was used to divide the long transcripts into smaller segments. We used the corresponding time information to segment the long audio files into smaller chunks using FFmpeg[1] and Python.

```
<?xml version="1.0" encoding="ISO-8859-1"?>
<!DOCTYPE Trans SYSTEM "trans-13.dtd">
<Trans scribe="(unknown)" audio_filename="digits1" version="2" version_date="031105">
<Episode>
<Section type="report" startTime="0" endTime="36.497">
<Turn startTime="0" endTime="36.497">
<Sync time="0"/>
sil
<Sync time="0.985"/>
five
<Sync time="1.735"/>
two
<Sync time="2.289"/>
four
<Sync time="2.852"/>
sp
<Sync time="3.445"/>
seven
<Sync time="4.098"/>
sp
<Sync time="4.258"/>
five
<Sync time="4.727"/>
nine
<Sync time="5.289"/>
sp
<Sync time="5.883"/>
one
<Sync time="6.414"/>
oh
<Event desc="error_wrong_word(s)" type="noise" extent="instantaneous"/>
<Sync time="6.914"/>
one
<Sync time="7.289"/>
sil
```

**FIGURE 3. EXAMPLE OF '.TRS' FILE IN THE PFSTAR DATASET.**
The content in this image was segmented into 'five two four', 'seven', 'five nine', and 'one oh one'.

The audio files from the PFSTAR dataset which were 30-70 seconds long were segmented into smaller audio chunks of 5-20 seconds in duration. This segmentation led to 12 hours of clean, usable PFSTAR data, having an audio length between 5-20 seconds. This 12 hours of data was further divided into 2 sets: PFS_10h with 10 hours of data (used in training) and PFS_test which contains 2 hours of PFSTAR data for inference. The final audio data has 16Khz sampling rate mono channel audio files in .wav format and transcriptions in .txt format.

## 3) CMUKIDS CLEANUP

CMU_Kids dataset also contains a lot of noisy and incomprehensible child speech. The transcriptions are provided in a '.trn' file format and audio files in a '.sph' format. The data was cleaned in a similar way to MyST by removing all the unrequired tags and non-textual information from the transcripts. For example, "they [begin_noise] kept a few [end_noise] butterflies in [noise]" was converted to "they kept a few butterflies in". A few more examples can be seen below:

o   [begin_noise] cages [end_noise] to lay more eggs [noise] [sil]
    -> <u>cages to lay more eggs</u>
o   a [begin_noise] blue butterfly [end_noise] /F L R UW/ [human_noise] flew by [human_noise] [human_noise]
    -> <u>a blue butterfly flew by</u>

The cleaned dataset contained all the audio files in '.wav' format and all transcribed speech in '.txt' format as needed for our training. The total amount of CMU_Kids dataset amounted to 9 hours which was used during inference only.

## C. DATASET USAGE

The dataset usage is mentioned in Table I. The table lists the datasets and their respective durations that were used for different purposes in the study. The 'Usage' column indicates whether the dataset was used for pretraining, finetuning, or inference. The 'Type' column specifies whether the dataset consists of child or adult speech.

TABLE I
DATASET DESCRIPTION FOR PRETRAINING, FINETUNING AND INFERENCE

| Usage | Dataset | Duration | Type |
|---|---|---|---|
| **Pretraining** [Unlabeled data] | MyST_complete | 393 hrs | Child |
| | Librispeech | 960 hrs | Adult |
| | Libri-light | 60k hrs | Adult |
| **Finetuning** [Labeled data] | MyST_10m | 10 mins | Child |
| | MyST_1h | 1 hr | Child |
| | MyST_10h | 10 hrs | Child |
| | MyST_55h | 55 hrs | Child |
| | PFS_10m | 10 mins | Child |
| | PFS_1h | 1 hr | Child |
| | PFS_10h | 10 hrs | Child |
| | LS_10m | 10 mins | Adult |
| | LS_100h | 100 hrs | Adult |
| | LS_960h | 960 hrs | Adult |
| **Inference** [Labeled data] | MyST_test | 10 hrs | Child |
| | PFS_test | 2 hrs | Child |
| | CMU_Kids | 9 hrs | Child |
| | LibriTTS 'dev-clean' | 8.9 hrs | Adult |

---

[1] FFmpeg: https://ffmpeg.org/

For pretraining, the complete MyST dataset, which consists of 393 hours of child speech, and the Librispeech and Libri-light datasets, which consist of 960 hours and 60,000 hours of adult speech, respectively, were used. No labels were used in pretraining.

The size of the finetuning datasets was chosen based on previous research in wav2vec2 [1] and to keep it consistent with the author's methodology. A similar distribution was maintained for finetuning with child speech datasets. The data was selected randomly to create subsets for finetuning. For finetuning, LS_10m, LS_100h, and LS_960h, which consist of 10 minutes, 100 hours, and 960 hours of Librispeech adult speech, respectively, was used. In addition, several subsets of MyST and PFSTAR datasets were used in finetuning as well, including MyST_10m, MyST_1h, MyST_10h, and MyST_55h, which consist of 10 minutes, 1 hour, 10 hours, and 55 hours of MyST child speech, respectively, and PFS_10m, PFS_1h, and PFS_10h, which consist of 10 minutes, 1 hour, and 10 hours of PFSTAR child speech, respectively.

For inference, the MyST_test, PFS_test, and CMU Kids datasets, which consist of 10 hours, 2 hours, and 9 hours of child speech, respectively, and the LibriTTS "dev-clean" dataset, which consists of 8.9 hours of adult speech, were used. Child speech subsets used for inference are not used in training.

## IV. CODEBASE AND EXPERIMENTS

### A. CODEBASE AND HYPERPARAMETERS
The wav2vec2 implementation provided by the fairseq[2] framework is used for our experiments. Hyperparameters were kept the same for both BASE and LARGE pretraining configurations as provided by the wav2vec2 authors. Finetuning configurations were also kept consistent with the finetuning dataset size used. Data cleaning and data processing scripts were created using FFmpeg and Python-based tools such as pydub and scipy. All the training checkpoints are made available on our GitHub page[3] and can be used directly with the model implementation from fairseq. See note[4] for more information on data cleaning scripts and dataset availability.

### B. EXPERIMENTS
Multiple pretraining and finetuning trainings were experimented with, depending on the dataset used (adult, child, or both). Experiments were divided into five groups, according to their configurations, as can be seen in Tables {II, III, IV, and V}. The five groups are called Group-A, Group-B, Group-C, Group-D and Group-E. For each of the groups, the ASR performance is measured in terms of Word Error Rate (WER) on different adult and child speech datasets. Child speech datasets used in inference include unseen MyST_10h, PFS_test and CMU_KIDS, and adult speech dataset include LibriTTS 'dev-clean'. These datasets remain the same for all groups during inference. To further differentiate between each experiment, 'Model ID' is used to make comparisons between different experiments.

#### 1) GROUP-A
For Group-A, the pretrained checkpoints provided by the wav2vec2 repository were used. Two configurations were used in Group-A training, namely BASE and LARGE. The BASE configuration includes 960 hours of Librispeech pretraining data and the LARGE configuration includes 60k hours of Librilight data, which is 60 times as much pretraining data as in the BASE configuration. The two configurations have different hyperparameters settings. For finetuning, each of the BASE and LARGE configurations were finetuned with 10 minutes, 100 hours, and 960 hours of Librispeech adult speech. Note that in Group-A, adult speech is being used for both pretraining and finetuning. Details of each model and their pretraining/finetuning datasets can be seen in Table II.

#### 2) GROUP-B
For Group-B, the pretrained model trained on adult speech is used and finetuned over different amounts of the MyST dataset. We use the BASE and LARGE configuration pretrained with 960 and 60k hours of adult speech (similar to Group-A) and finetune it over 10 minutes, 1 hour, 10 hours, and 55 hours of MyST child speech data. Therefore, Group-B contains adult speech data in pretraining and child speech data in finetuning steps. Refer to Table III for more details.

#### 3) GROUP-C
In Group-C, the Librispeech and MyST datasets having 960 hours of adult speech and 393 hours of child speech data, respectively, was used for pretraining. The model is then finetuned over different amounts of the MyST dataset (similar to Group-B). Therefore, Group-C contains both adult and child speech data in pretraining and only child speech in finetuning. The BASE configuration was used for this experiment, and we did not experiment with the LARGE configuration. This will be explained in the Results section. See Table III for more details about the experimental groups.

#### 4) GROUP-D
Group-D is very similar to Group-B as the PFSTAR dataset is used for finetuning instead of the MyST dataset. We use the BASE and LARGE configuration pretrained with 960 and 60k hours of adult speech and finetune it with 10 minutes, 1 hour and 10 hours of PFSTAR child speech dataset. PFSTAR is used as it's a British English child to other research material. For access to respectively cleaner versions of datasets used in this paper, researchers can buy their own license for the original datasets (where required), and on providing proof of that license, can get access to our 'clean' versions.

speech dataset and will have different acoustic properties than the American English MyST child speech dataset. More on this in Section-V. Group-D experiments and datasets used are mentioned in Table IV.

### 5) GROUP-E

Group-E uses a mix of different datasets in the finetuning. We use the same BASE and LARGE configurations pretrained with 960 and 60k hours of adult speech as in the previous groups. For finetuning, a mix of the MyST_55h, PFS_10h, and LS_960 datasets was used. Finetuning data included LS_960h+MyST_55h, LS_960h+PFS_10h, MyST_55h+PFS_10h and LS_960h+MyST_55h+ PFS_10h. These experiments were performed to see the correlation in WER with respect to different finetuning datasets used. Experimental details are mentioned in Table V.

All these experiments were designed to see the effect of different pretraining and finetuning configurations and different amounts of adult and child speech datasets used in these configurations. We did not train any independent pretrained model with child speech data alone as larger amounts of child speech data would be required to learn any meaningful speech representations from child speech (more in Section V.)

## V. RESULTS AND DISCUSSION

### A. MAIN RESULTS FROM THE GROUP EXPERIMENTS

The results are presented by calculating WER, a metric typically used for evaluating ASR systems. The WER uses edit distance to calculate differences between the reference transcript and model-generated output.

### 1) GROUP-A

For Group-A training (see Table I for details), Model 6, pretrained on 60k hours of adult speech and finetuned on 960 hours of adult speech, gave the best results with a WER of 12.50 on MyST_test, 8.56 on PFS_test, 14.85 on CMU_KIDS and 3.28 on dev-clean. All models have a similar pattern of decreasing WER with an increase in the size of the finetuning dataset. It can also be observed that there is not a large difference in WER for both BASE and LARGE configurations when changing dataset size from LS_100h to LS_960h. There is a 10x difference in dataset size but the change in WER is not more than 3% on test datasets. Therefore, 100 hours of finetuning data can be considered an ideal amount of adult speech data to be used with child speech validation in low-data scenarios.

The LARGE configuration, having approximately 60x difference in pretraining data, performs better. The models that used the LARGE configuration {4, 5, 6} generally have lower WER values on the test datasets compared to the models that used the BASE configuration {1, 2, 3}. This suggests that the LARGE configuration may be more effective at improving the performance of the model, implying that more pretraining data leads to better learnable representations. The average WER (see Table-II) is observed to be 21.57, 19.07, 22.81, 7.82 for BASE and 17.37, 15.78, 20.13, 7.47 for the LARGE configuration when inferred over MyST_test, PFS_test, CMU_Kids and dev_clean datasets, respectively. Between the BASE and LARGE configurations, there is a similar decrease in WER over MyST_test, CMU_Kids and PFS_test datasets, implying that adult speech finetuning had a similar effect on the performance of the ASR model when used on child speech datasets.

### 2) GROUP-B

For experiments in Group-B (see Table-II for more details), a similar trend of decreasing WER can be observed with an increase in the amount of finetuning data. Models 10 and 14 with BASE and LARGE configurations respectively, and finetuned on 55 hours of MyST child speech gave the best results on all four test datasets. A large decrease in WER can be observed not just with the inference on MyST but also with the other child speech datasets. All the models in Group-B, finetuned with different amounts of MyST data, attained lower WERs on the child speech datasets in comparison with Group-A experiments, which had more training data. This could imply that a model finetuned with child speech data can achieve meaningful improvements for all the different types of different child speech datasets. Interestingly, child speech finetuning not only lead to an increase in the performance of child speech inference, but also had a significant effect on adult speech inference.

When comparing the best models from Group-A (Table-I) and Group-B (Table-II), i.e., models 6 and 14 having a similar pretraining configuration, model 14 finetuned with 55 hours of child speech achieves better a WER of 7.51 on MyST_test than model 6 (with WER of 12.50) finetuned on 960 hours of adult speech. 'dev-clean' adult speech dataset also reached a 6.53 WER when trained with MyST_55h, which is just double the size of WER when finetuned with 960 hours of adult speech (i.e., 3.28). However, a performance decline can be seen in PFSTAR_test inference as the WER increased from 8.56 in model 6 to 12.46 in model 14, and similarly in CMU_KIDS, where WER increased from 14.85 in model 6 to 15.25 in model 14. This could imply that models finetuned with adult datasets perform similarly in inference on different child speech datasets with different distributions (as in Group-A), but when finetuned with child speech, performance only improves during inference on child speech with the same or similar distribution as the finetuning set. Model 9 finetuned with MyST_10h outperforms model 3 finetuned with LS_960h during inference on MyST_test, although both models have the same BASE configuration. This indicates that 10 hours of child speech dataset is enough to outperform the SOTA models trained with 960 hours of adult speech (for a specific/known child speech distribution).

Comparing the BASE and LARGE experiments in Group-B, average WERs of 17.29, 29.15, 23.34, 13.51 and 17.08, 27.02, 22.91, 8.62 are observed for the BASE and LARGE configurations respectively, when inferred over MyST_test, PFS_test, CMU_Kids, and dev_clean datasets, respectively. There is a slight decrease in WER for all child speech datasets between the BASE and LARGE models. The average difference in WER for child speech datasets between the BASE and LARGE experiments is less than 2%, This indicates that the 60x increase in pretraining adult speech data did not lead to a major increase in the performance of ASR models on child speech when pretrained with adult speech and finetuned with child speech. Therefore, the BASE configuration can be considered ideal for child speech finetuning in a low-data scenario. It can also be observed that there is a huge increase in WER of adult speech from BASE to LARGE configurations. On a more detailed inspection, it was observed that this change only happens when finetuned with 10 minutes of child speech (see model 7 and model 11). If only 10 minutes of labeled child speech is available, then a larger pretraining dataset is detrimental because of the huge mismatch between pretraining and finetuning data distributions. In all other cases, finetuning on the larger dataset brings improvements. Hence, the LARGE pretraining model needs enough child data during finetuning to be able to cope with the data distribution mismatch.

### 3) GROUP-C

Group-C experiments were designed similarly to Group-B (see Table-III for more details). Observing the Group-C experiments, a similar pattern of a decrease in WER with an increase in finetuning data can be seen during the inference on the test data. In Group-C, the MyST_Complete set, which contains 343 hours of MyST child speech is included in the pretraining along with 960 hours of the Librispeech adult speech dataset. When comparing the BASE models {7,8,9,10} and {15,16,17,18} from Group-B (see Table-II) and Group-C (see Table-III) respectively, it can be observed that by including the MyST dataset in the pretraining, the WERs on all test sets increased in Group-C as compared to the corresponding WERs in Group-B. The average WER between Group-B and Group-C BASE experiments is observed to be 17.29, 29.15, 23.34 and 13.51 in Group-B BASE and 19.39, 35.67, 28.7 and 21.53 in Group-C BASE experiments when inferred over MyST_test, PFS_test, CMU_Kids and dev_clean datasets respectively. There is a large increase in WER from Group-B to Group-C. This can be attributed to the MyST_Complete corpus containing a lot of noise and non-linguistic child speech data, and it will be difficult to learn meaningful representations from such 'noisy' data and without any labels.

This observation could imply that a model pretrained with only adult speech data can learn better features than the model pretrained with both adult and (noisy) child speech data in wav2vec2 SSL; however, more investigation is required to determine why and how the pretraining of child speech data affects these models. This would require a much cleaner, noise-free and larger child speech dataset. Therefore, due to this limitation, the child speech dataset in pretraining was not considered for the Group-D and Group-E experiments.

TABLE II
GROUP-A: WER FOR DIFFERENT PRETRAINING (ADULT SPEECH DATASETS) AND FINETUNING (ADULT SPEECH DATASET) EXPERIMENTS ON THE MYST, PF-STAR, CMU KIDS AND LIBRITTS 'DEV-CLEAN' DATASETS.

| Group | Model ID | Pretraining Model Configuration | Pretraining dataset | Finetuning dataset | WER MyST_test | WER PFS_test | WER CMU_KIDS | WER dev_clean |
|---|---|---|---|---|---|---|---|---|
| GROUP - A | 1 | BASE | Librispeech | LS_10m | 31.48 | 30.05 | 33.38 | 15.90 |
| | 2 | | | LS_100h | 17.82 | 15.96 | 18.73 | 4.16 |
| | 3 | | | LS_960h | **15.41** | **11.20** | **16.33** | **3.40** |
| | - | *Average (Group – A, BASE)* | | | *21.57* | *19.07* | *22.81* | *7.82* |
| | 4 | LARGE | Librilight | LS_10m | 26.47 | 27.14 | 29.37 | 15.35 |
| | 5 | | | LS_100h | 13.15 | 11.63 | 16.18 | 3.79 |
| | 6 | | | LS_960h | **12.50** | **8.56** | **14.85** | **3.28** |
| | - | *Average (Group – A, LARGE)* | | | 17.37 | 15.78 | 20.13 | 7.47 |

TABLE III
GROUP-B AND GROUP-C: WER FOR DIFFERENT PRETRAINING (ADULT AND CHILD SPEECH DATASETS) AND FINETUNING (MYST CHILD SPEECH DATASET) COMBINATIONS ON THE MYST, PF-STAR, CMU KIDS AND LIBRITTS 'DEV-CLEAN' DATASETS.

| Group | Model ID | Pretraining Model Configuration | Pretraining dataset | Finetuning dataset | WER MyST_test | WER PFS_test | WER CMU_KIDS | WER dev_clean |
|---|---|---|---|---|---|---|---|---|
| | 7 | BASE | Librispeech | MyST_10m | 28.84 | 41.34 | 34.18 | 21.45 |

| Group | Model ID | Pretraining Model Configuration | Pretraining dataset | Finetuning dataset | WER MyST_test | WER PFS_test | WER CMUKIDS | WER dev_clean |
|---|---|---|---|---|---|---|---|---|
| GROUP-B | 8 | | | MyST_1h | 18.75 | 31.84 | 23.13 | 13.91 |
| | 9 | | | MyST_10h | 13.46 | 28.68 | 19.59 | 10.94 |
| | 10 | | | MyST_55h | **8.13** | **14.77** | **16.47** | **7.72** |
| | - | *Average (Group – B, BASE)* | | | *17.29* | *29.16* | *23.34* | *13.51* |
| | 11 | LARGE | Librilight | MyST_10m | 33.01 | 44.36 | 39.91 | 46.45 |
| | 12 | | | MyST_1h | 14.91 | 26.21 | 18.74 | 11.59 |
| | 13 | | | MyST_10h | 12.92 | 25.05 | 17.72 | 10.04 |
| | 14 | | | MyST_55h | **7.51** | **12.46** | **15.25** | **6.43** |
| | - | *Average (Group – B, LARGE)* | | | *17.08* | *27.02* | *22.91* | *18.62* |
| GROUP-C | 15 | BASE | Librispeech MyST_Complete | MyST_10m | 29.16 | 45.71 | 37.56 | 35.39 |
| | 16 | | | MyST_1h | 21.89 | 38.53 | 29.03 | 20.45 |
| | 17 | | | MyST_10h | 16.18 | 32.95 | 25.06 | 16.83 |
| | 18 | | | MyST_55h | **10.34** | **25.47** | **23.15** | **13.48** |
| | - | *Average (Group – C, BASE)* | | | *19.39* | *35.67* | *28.7* | *21.53* |

TABLE IV

GROUP-D: WER for different Pretraining (adult speech datasets) and Finetuning (PFstar child speech dataset) combinations on the MYST, PF-STAR, CMU KIDS and LIBRITTS 'dev-clean' datasets.

| Group | Model ID | Pretraining Model Configuration | Pretraining dataset | Finetuning dataset | WER MyST_test | WER PFS_test | WER CMUKIDS | WER dev_clean |
|---|---|---|---|---|---|---|---|---|
| GROUP-D | 19 | BASE | Librispeech | PFS_10m | 35.91 | 16.43 | 33.53 | 30.43 |
| | 20 | | | PFS_1h | 33.52 | 7.36 | 29.55 | 16.61 |
| | 21 | | | PFS_10h | **31.86** | **3.48** | **27.49** | **13.95** |
| | - | *Average (Group – D, BASE)* | | | *33.76* | *9.09* | *30.19* | *20.33* |
| | 22 | LARGE | Librilight | PFS_10m | 37.10 | 16.78 | 35.13 | 23.85 |
| | 23 | | | PFS_1h | 30.81 | 14.19 | 28.54 | 21.89 |
| | 24 | | | PFS_10h | **27.17** | **3.50** | **21.35** | **11.60** |
| | - | *Average (Group – D, LARGE)* | | | *31.69* | *11.49* | *28.34* | *19.11* |

TABLE V

GROUP-E: WER for different Pretraining (adult datasets) and Finetuning (adult and child speech datasets) combinations on the MYST, PF-STAR, CMU KIDS and LIBRITTS 'dev-clean' datasets.

| Group | Model ID | Pretraining Model Configuration | Pretraining dataset | Finetuning dataset | WER MyST_test | WER PFS_test | WER CMUKIDS | WER dev_clean |
|---|---|---|---|---|---|---|---|---|
| GROUP-E | 25 | BASE | Librispeech | LS_960h, MyST_55h | 8.18 | 12.17 | 14.12 | 1.24 |
| | 26 | | | LS_960h, PFS_10h | 15.42 | 3.74 | 15.31 | 1.41 |
| | 27 | | | MyST_55h, PFS_10h | **7.94** | **2.91** | 15.97 | 7.64 |
| | 28 | | | LS_960h, MyST_55h, PFS_10h | 8.13 | 3.12 | **13.76** | **1.20** |
| | - | *Average (Group – E, BASE)* | | | *9.91* | *5.48* | *14.79* | *2.87* |
| | 29 | LARGE | Librilight | LS_960h, MyST_55h | 8.06 | 9.31 | 13.20 | 1.34 |
| | 30 | | | LS_960h, PFS_10h | 13.18 | 3.17 | 13.19 | **1.32** |
| | 31 | | | MyST_55h, PFS_10h | **7.42** | **2.99** | 14.18 | 5.79 |
| | 32 | | | LS_960h, MyST_55h, PFS_10h | 8.17 | 3.33 | **12.77** | 1.40 |
| | - | *Average (Group – E, LARGE)* | | | *9.2* | *4.7* | *13.33* | *2.4* |

### 4) GROUP-D

Group-D experiments (see Table-IV) also contain BASE and LARGE configurations, where Librispeech and Librilight datasets are used for pretraining, and PFS_10m, PFS_1h, and PFS_10h are used for finetuning. The model's performance appears to improve as the size of the finetuning dataset increases, with the 'PFS_10h' dataset yielding the lowest WER values on all test groups. Comparing Group-B (from Table-II) and Group-D (from Table-III) having different children finetuning datasets, it can also be seen that the WER decreased significantly for the PFS_test but hardly for MyST_test and CMU_Kids. This can be attributed to the accents in the datasets. Both MyST and CMU_Kids datasets contain American English child speech, while the PFSTAR dataset contains British English child speech. Therefore, a similar pattern of change in WER can be seen in the inference on MyST and CMU_Kids. This could also be due to differences in the acoustic characteristics of the audio between theMyST/CMU_Kids and PFSTAR datasets. Model 24 with LARGE pretraining configuration and finetuned with PFS_10h gave the lowest WER in this group for all the inference data.

To extend on the previous observation, we can take a look at the average WER between models {7,8,9,11,12,13} in Group-B (from Table-II) and models {19,20,21,22,23,24} in Group-D (from Table-IV), which have the same amount of MyST and PFSTAR finetuning data. The average WER for MyST_test increased from 20.32 in Group-B to 32.72 in Group-D, PFS_test decreased from 32.91 in Group-B to 10.29 in Group-D and CMU_Kids also increased from 25.55 in Group-B to 29.26 in Group-D. It can be observed that when finetuning with MyST subsets (American English), the WER for MyST_test and CMU_Kids decreased, both being American English datasets, but WER increased for PFS_test (British English) and vice versa. Hence, we can say that properties like the dialect of speaking, accent, and acoustic properties in different child speech datasets can also affect the performance of the ASR model. This also implies that the trained model would have significant improvement during inference on test datasets (having similar distribution to the finetuning dataset) as opposed to other unseen datasets. For inference over adult speech, the average WER on LibriTTS dev-clean was observed to be 19.06 for MyST finetuning in models {7,8,9,11,12,13} (see Table-II) and 19.72 for PFSTAR finetuning in models {19,20,21,22,23,24} (see Table-IV). The difference in WER is less than 1%. From this observation, it can be concluded that different child datasets finetuning have similar effects on accuracy when inferred with adult speech.

If we compare model 6 (from Table-II) and model 24 (from Table-IV) for the PFS_test, having the same LARGE configuration, 10 hours of PFSTAR finetuning was able to outperform model 6 with 960 hours of adult speech finetuning. Hence, 10 hours can be considered an optimum amount for finetuning child speech in a low-data scenario.

It can also be observed that the LARGE configuration has slightly higher WERs than the BASE configurations' WERs on PFS_test but have a pattern of decreasing WER in MyST_test, CMU_Kids and dev-clean. The average WERs for MyST_test, CMU_KIDS and dev-clean decreased from 33.76, 30.19 and 20.33 in the BASE configuration to 31.69, 28.34, and 19.11, respectively, in the LARGE configuration, while the average WER for PFS_test increased from 9.09 in BASE to 11.49 in LARGE configuration. This also implies that MyST_test and CMU_Kids test with American English child speech have relatively similar acoustic properties to pretraining adult speech datasets than the PFS_test British English dataset. Also, this behavior was only observed in Group-D (with PFSTAR finetuning). It implies that BASE configuration with 960 hours is better for finetuning with the PFSTAR dataset as compared to LARGE configuration. It is also observed that 1 hour of PFSTAR finetuning (model 23) is not enough when finetuning with a larger pretraining dataset because of a huge mismatch between pretraining and finetuning data distributions.

### 5) GROUP-E

Group-E results (from Table-V) included BASE and LARGE configurations pretrained on Librispeech and Librilight datasets and finetuned with a combination of two or more finetuning datasets. In Group-E, finetuning datasets which achieved the best WER in all the previous groups were used. LS_960h, PFS_10h and MyST_55h were used in different combinations as these are the datasets with the most amount of pretraining data (in hours) and gave the best WER in their different use cases in previous experiments. Group-E models outperformed all the previous models in terms of WER. The best WERs were achieved for all the inference datasets used in Group-E experiments. Our Group-E results also outperformed all the previously reported results on these inference datasets including both supervised and SSL approaches.

The BASE and LARGE configurations in Group-E, having a 60x difference in pretraining datasets, do not show a lot of difference in their WERs on inference datasets. The average WER for the BASE experiments over MyST_test, PFS_test, CMU_Kids and dev_clean are 9.91, 5.4, 14.79 and 2.87, respectively, (see Table-V), whereas the average WERs for the LARGE experiments for the same inference datasets are 9.2, 4.7, 13.33, and 2.4 (from Table-V). The overall absolute difference in WER for the BASE and LARGE configurations is 0.84. The Librilight dataset only gave a WER improvement of 0.84 over Librispeech. From this observation, it can be concluded that the performance remained the same for both BASE and LARGE configurations when cross-domain combinations of datasets were used in finetuning. Therefore, the BASE configuration pretrained with 960 hours of adult speech is ideal for finetuning with combinations of child speech finetuning in a low-data scenario.

TABLE VI
OUR RESULTS VS PREVIOUS SOTA RESULTS ON THE MYST, PF-STAR, AND CMU_KIDS DATASETS.

| SOTA Papers | Method Type | Training Data (hrs) | Inference data (hrs) | WER MyST | WER PFSTAR | WER CMU_Kids |
|---|---|---|---|---|---|---|
| This work {Model 27, 31, 32} | Self-Supervised | see Table I | see Table I | **7.42** | **2.91** | **12.77** |
| TDNN-F + Augmentation [26] | Supervised | 6.34 | 2.76 | - | - | 16.01 |
| Hybrid HMM-DNN Transfer Learning [24] | Supervised | 6.26 | 2.45 | - | - | 19.33 |
| DRAFT [33]:<br>○ WAV2VEC2<br>○ HuBERT | Self-Supervised | 197 | 13 | 16.70<br>16.53 | - | - |
| Transformer + CTC + Greedy [25] | Supervised | 197 | 13 | 16.01 | - | - |
| W2v2 + source-filter warping + LM [31] | Self-Supervised | 11.2 | 2.5 | | 4.86 | |

Each of the models in Group-E outperforms all the experiments in groups {A, B, C and D} on all four inference datasets. There is a decrease in WER for both adult and child speech inference datasets. The best WER of 1.20 was achieved on dev-clean, which outperforms all the experiments in Group-A when only adult speech was used in finetuning. This implies that introducing child speech in finetuning also helps in improving the performance of ASR on adult speech.

While the results of this study provide a comprehensive analysis of the effectiveness of different finetuning techniques and configurations on child speech recognition, further research can be conducted to draw additional conclusions by comparing different groups of experiments and data used. This study provides a baseline for future studies. Further improvements can be made in ASR performance on child speech by using additional methods like the new speech augmentations, generative child speech, language models etc.

### B. COMPARISON WITH THE PREVIOUS SOTA APPROACHES

Our results using the SSL approach can outperform all the previously reported results on the same datasets. This can be seen in Table VI. Our trained models have better WER for MyST, PFSTAR and CMU_KIDS datasets, indicating better child ASR performance.

**MyST:** We achieved the best WER of **7.42** on the MyST dataset which outperforms the previously reported SOTA WER of 16.01, that was achieved with a supervised transformer model [25]. Our result also outperforms previously reported results with SSL having a WER of 16.70 using the DRAFT wav2vec2 SSL approach [33] and 16.53 using the DRAFT HuBERT learning method [33] on the same dataset. We only compare with the best results from these papers [25], [33] as they contain many different comparisons and findings with different training approaches.

**PFSTAR:** A WER of **2.91** was achieved on the PFSTAR dataset, reaching human level performance and outperforming previously reported WER of 4.86 with wav2vec2 using source-filter warping and language model [31].

**CMU_Kids:** Our results also outperform all the previously reported results on the unseen CMU_Kids dataset without using any CMU_kids dataset in any training. We achieved a WER of **12.77** on CMU_Kids outperforming the previously reported results with the supervised learning approaches. The authors in [26] used TDNN-F and augmentation to achieve the previously best WER of 16.01 on the CMU_Kids dataset. Authors in [24] used a similar method with the Hybrid HMM-DNN multilingual transfer learning approach and achieved a WER of 19.33 on the CMU_Kids dataset.

### VI. CONCLUSION

In this work, the wav2vec2 self-supervised training approach is used with different pretraining and finetuning datasets to improve child speech recognition. A combination of adult and child speech datasets was used in pretraining and finetuning to find the ideal data requirements for improving child speech recognition. Experiments were designed to see the performance on the in-domain MyST and PFSTAR datasets, and the out-of-domain CMUKIDS dataset. LibriTTS dev-clean was also used as a clean adult speech test dataset to compare the performance on adult speech.

The best results were obtained from Group-E experiments, where the model was finetuned on combinations of adult and child speech datasets with the best WER of 7.42 on MyST_test, 2.91 on PFS_test, 12.77 on CMU_KIDS, and 1.20 on dev-clean, outperforming the previous SOTA results on the same datasets. Our results indicate that by using self-supervised adult speech representations and finetuning over child corpora, we can effectively overcome the issue of data scarcity in children's speech recognition.

A model pretrained with adult speech data can learn the best features as compared to a model including both adult and child speech in pretraining. Adding the MyST child speech dataset in pretraining reduced the performance of the ASR model on all test datasets, which is attributed to the 'noisy' nature of the MyST dataset. It would be interesting to include a clean child speech dataset in the pretraining and compare results on that. The LARGE configuration, which was pretrained with 60 times more data than the BASE configuration, performed 3% better overall when finetuned with a single dataset. Both the BASE and LARGE pretraining gave an almost similar performance on test datasets when finetuned with combinations of datasets. Therefore, the BASE configuration can be used effectively in a low-data scenario.

Finetuning with adult speech data generally leads to a decrease in WER as the amount of finetuning data increases, regardless of the specific child speech dataset used in inference and shows similar performance on all the different child speech test datasets. And similarly, models finetuned with different child datasets show similar performance on adult speech test dataset used during inference. Fine-tuning with child speech data also achieved improvements on all different types of child speech test datasets and shows significant improvements on adult speech inference as well. Even a small amount of child speech data as low as 1 hour during finetuning can improve performance.

100 hours of adult speech finetuning data offer a practical trade-off between computational effort for training and the final performance level to be used with child speech validation in low data scenario situations. Low resource finetuning with 10 hours of child speech led to an increase in performance when compared to SOTA wav2vec2 models finetuned with 960 hours of adult speech. For a model finetuned with single or multiple child/adult speech data, WER increases over the dataset with similar distribution as finetuning dataset. Hence, it can be difficult to generalize on child speech datasets for ASR use cases.

It was also observed that American English datasets such as MyST and CMU_kids child speech datasets have many similar acoustic properties to the Librispeech adult speech dataset as compared to the British English PFSTAR child speech dataset.

For future work, we intend to use the model with the best WER to transcribe more child speech data from the unlabeled MyST dataset. It would also be interesting to investigate the potential of generative data augmentation models [43] to generate synthetic child speech for providing a wider variety of child speech for pretraining and finetuning experiments. We also intend to conduct more experiments by incorporating other child and adult speech datasets using this SSL approach. Additionally, we will explore using external language models for the validation of child speech in the SSL domain.

## ACKNOWLEDGMENT

The authors would like to acknowledge experts from Xperi Ireland: Gabriel Costache, Zoran Fejzo, George Sterpu and the rest of the team members for providing their expertise and feedback throughout.

## REFERENCES

[1] A. Baevski, H. Zhou, A. Mohamed, and M. Auli, "wav2vec 2.0: A Framework for Self-Supervised Learning of Speech Representations", Advances in neural information processing systems, 33, pp. 12449-12460, 2020.

[2] Amodei, Dario, et al. "Deep speech 2: End-to-end speech recognition in english and mandarin." *International conference on machine learning.* PMLR, 2016.

[3] Hannun, Awni, et al. "Deep speech: Scaling up end-to-end speech recognition." *arXiv preprint arXiv:1412.5567* (2014).

[4] Kriman, Samuel, et al. "Quartznet: Deep automatic speech recognition with 1d time-channel separable convolutions." *ICASSP 2020-2020 IEEE International Conference on Acoustics, Speech and Signal Processing (ICASSP).* IEEE, 2020. Available: https://github.com/NVIDIA/NeMo

[5] A. B. Nassif, I. Shahin, I. Attili, M. Azzeh, and K. Shaalan, "Speech Recognition Using Deep Neural Networks: A Systematic Review," *IEEE Access*, vol. 7, pp. 19143–19165, 2019, doi: 10.1109/ACCESS.2019.2896880.

[6] Han, Wei, et al. "Contextnet: Improving convolutional neural networks for automatic speech recognition with global context." *arXiv preprint arXiv:2005.03191* (2020).

[7] Gulati, Anmol, et al. "Conformer: Convolution-augmented transformer for speech recognition." *arXiv preprint arXiv:2005.08100* (2020).

[8] Lee, Sungbok, Alexandros Potamianos, and Shrikanth Narayanan. "Acoustics of children's speech: Developmental changes of temporal and spectral parameters." *The Journal of the Acoustical Society of America* 105.3 (1999): 1455-1468.

[9] Lee, Sungbok, Alexandros Potamianos, and Shrikanth Narayanan. "Analysis of children's speech: Duration, pitch and formants." *Fifth European Conference on Speech Communication and Technology.* 1997.

[10] F. Claus, H. Gamboa Rosales, R. Petrick, H.-U. Hain, and R. Hoffmann, "A survey about databases of children's speech." *INTERSPEECH.* 2013.

[11] S. Shahnawazuddin, N. Adiga, H. K. Kathania, and B. T. Sai, "Creating speaker independent ASR system through prosody modification based data augmentation," *Pattern Recognit Lett,* vol. 131, pp. 213–218, Mar. 2020, doi: 10.1016/j.patrec.2019.12.019.

[12] W. Wang, Z. Zhou, Y. Lu, H. Wang, C. Du, and Y. Qian, "Towards data selection on TTS data for children's speech recognition," *ICASSP, IEEE International Conference on Acoustics, Speech and Signal Processing - Proceedings,* vol. 2021-June, pp. 6888–6892, 2021, doi: 10.1109/ICASSP39728.2021.9413930.

[13] V. Kadyan, H. Kathania, P. Govil, and M. Kurimo, "Synthesis Speech Based Data Augmentation for Low Resource Children ASR," *Lecture Notes in Computer Science (including subseries Lecture Notes in Artificial Intelligence and Lecture Notes in Bioinformatics),* vol. 12997 LNAI, pp. 317–326, 2021, doi: 10.1007/978-3-030-87802-3_29.

[14] S. Shahnawazuddin, N. Adiga, K. Kumar, A. Poddar, and W. Ahmad, "Voice Conversion Based Data Augmentation to Improve Children's Speech Recognition in Limited Data Scenario," 2020, doi: 10.21437/Interspeech.2020-1112.

[15] D. K. Singh, P. P. Amin, H. B. Sailor, and H. A. Patil, "Data Augmentation Using CycleGAN for End-to-End Children ASR," *European Signal Processing Conference,* vol. 2021-August, pp. 511–515, 2021, doi: 10.23919/EUSIPCO54536.2021.9616228.

[16] N. Jia, C. Zheng, and W. Sun, "Speech Synthesis of Children's Reading Based on CycleGAN Model," *J Phys Conf Ser,* vol. 1607, p. 12046, Aug. 2020, doi: 10.1088/1742-6596/1607/1/012046.

[17]    Serizel, Romain, and Diego Giuliani. "Vocal tract length normalisation approaches to DNN-based children's and adults' speech recognition." *2014 IEEE Spoken Language Technology Workshop (SLT)*. IEEE, 2014.

[18]    G. Yeung, R. Fan, and A. Alwan, "Fundamental Frequency Feature Normalization and Data Augmentation for Child Speech Recognition," in *ICASSP 2021 - 2021 IEEE International Conference on Acoustics, Speech and Signal Processing (ICASSP)*, 2021, pp. 6993–6997. doi: 10.1109/ICASSP39728.2021.9413801.

[19]    J. Fainberg, P. Bell, M. Lincoln, and S. Renals, "Improving children's speech recognition through out-of-domain data augmentation," *Proceedings of the Annual Conference of the International Speech Communication Association, INTERSPEECH*, vol. 08-12-September-2016, pp. 1598–1602, 2016, doi: 10.21437/INTERSPEECH.2016-1348.

[20]    G. Chen *et al.*, "Data Augmentation For Children's Speech Recognition -- The 'Ethiopian' System For The SLT 2021 Children Speech Recognition Challenge," Nov. 2020, doi: 10.48550/arxiv.2011.04547.

[21]    Park, Daniel S., et al. "Specaugment: A simple data augmentation method for automatic speech recognition." *arXiv preprint arXiv:1904.08779* (2019).

[22]    V. P. Singh, H. Sailor, S. Bhattacharya, and A. Pandey, "Spectral Modification Based Data Augmentation For Improving End-to-End ASR For Children's Speech".

[23]    P. Gurunath Shivakumar and P. Georgiou, "Transfer learning from adult to children for speech recognition: Evaluation, analysis and recommendations," *Comput Speech Lang*, vol. 63, p. 101077, Sep. 2020, doi: 10.1016/J.CSL.2020.101077.

[24]    T. Rolland, A. Abad, C. Cucchiarini, and H. Strik, "Multilingual Transfer Learning for Children Automatic Speech Recognition," in *Proceedings of the Thirteenth Language Resources and Evaluation Conference*, Jun. 2022, pp. 7314–7320. [Online]. Available: https://aclanthology.org/2022.lrec-1.795

[25]    Shivakumar, Prashanth Gurunath, and Shrikanth Narayanan. "End-to-end neural systems for automatic children speech recognition: An empirical study." *Computer Speech & Language* 72 (2022): 101289.

[26]    F. Wu, L. Paola Garcia, D. Povey, and S. Khudanpur, "Advances in Automatic Speech Recognition for Child Speech Using Factored Time Delay Neural Network," 2019, doi: 10.21437/Interspeech.2019-2980.

[27]    Baevski, Alexei, et al. "Unsupervised speech recognition." *Advances in Neural Information Processing Systems* 34 (2021): 27826-27839. Available: https://github.com/pytorch/fairseq/tree/

[28]    G. Xu, S. Yang, L. Ma, C. Li, and Z. Wu, "The TAL system for the INTERSPEECH2021 Shared Task on Automatic Speech Recognition for Non-Native Childrens Speech," 2021, doi: 10.21437/Interspeech.2021-1104.

[29]    Yu, Fan, et al. "The SLT 2021 children speech recognition challenge: Open datasets, rules and baselines." *2021 IEEE Spoken Language Technology Workshop (SLT)*. IEEE, 2021.

[30]    Fan, Ruchao, Amber Afshan, and Abeer Alwan. "Bi-apc: Bidirectional autoregressive predictive coding for unsupervised pre-training and its application to children's asr." *ICASSP 2021-2021 IEEE International Conference on Acoustics, Speech and Signal Processing (ICASSP)*. IEEE, 2021.

[31]    J. Thienpondt and K. Demuynck, "Transfer Learning for Robust Low-Resource Children's Speech ASR with Transformers and Source-Filter Warping", 2022.

[32]    R. Fan, Y. Zhu, J. Wang, and A. Alwan, "Towards Better Domain Adaptation for Self-Supervised Models: A Case Study of Child ASR," *IEEE J Sel Top Signal Process*, vol. 16, no. 6, pp. 1242–1252, 2022, doi: 10.1109/JSTSP.2022.3200910.

[33]    R. Fan and A. Alwan, "DRAFT: A Novel Framework to Reduce Domain Shifting in Self-supervised Learning and Its Application to Children's ASR," 2022, doi: 10.21437/Interspeech.2022-11128.

[34]    E. Jang, G. Brain, S. Gu, and B. Poole, "Categorical reparameterization with gumbel-softmax," *ICLR* , 2017.

[35]    A. Graves, A. Ch, S. Fernández, F. Gomez, J. Schmidhuber, and J. Ch, "Connectionist Temporal Classification: Labelling Unsegmented Sequence Data with Recurrent Neural Networks", *Proceedings of the 23rd international conference on Machine learning*. 2006.

[36]    D. P. Kingma and J. Lei Ba, "ADAM: A METHOD FOR STOCHASTIC OPTIMIZATION," *ICLR*, 2015.

[37]    W. Wayne, R. Cole, and S. Pradhan, "My Science Tutor and the MyST Corpus," no. April, 2019, [Online]. Available: https://boulderlearning.com/request-the-myst-corpus/

[38]    M. Eskenazi, J. Mostow, and D. Graff, "The CMU kids speech corpus," *Corpus of children's read speech digitized and transcribed on two CD-ROMs, with assistance from Multicom Research and David Graff. Published by the Linguistic Data Consortium, University of Pennsylvania*, 1997.

[39]    Russell, Martin. "The pf-star british english childrens speech corpus." *The Speech Ark Limited* (2006).

[40]    Kahn, Jacob, et al. "Libri-light: A benchmark for asr with limited or no supervision." *ICASSP 2020-2020 IEEE International Conference on Acoustics, Speech and Signal Processing (ICASSP)*. IEEE, 2020.

[41]    H. Zen *et al.*, "LibriTTS: A corpus derived from libri speech for text-to-speech," *arXiv*. 2019. doi: 10.21437/interspeech.2019-2441.

[42]    V. Panayotov, G. Chen, D. Povey, and S. Khudanpur, "Librispeech: An ASR corpus based on public domain audio books," in *2015 IEEE International Conference on Acoustics, Speech and Signal Processing (ICASSP)*, Apr. 2015, pp. 5206–5210. doi: 10.1109/ICASSP.2015.7178964.

[43]    R. Jain, M. Y. Yiwere, D. Bigioi, P. Corcoran, and H. Cucu, "A Text-to-Speech Pipeline, Evaluation Methodology, and Initial Fine-Tuning Results for Child Speech Synthesis," *IEEE Access*, vol. 10, pp. 47628–47642, 2022, doi: 10.1109/ACCESS.2022.3170836.